\documentstyle[preprint,aps,psfig,subfigure]{revtex} 
\newcommand{\be}{\begin{equation}} 
\newcommand{\ee}{\end{equation}}
\newcommand{\ba}{\begin{eqnarray}} 
\newcommand{\ea}{\end{eqnarray}}
\newcommand{\no}{\nonumber \\} 
\newcommand{\bb}{\bibitem}
\newcommand{\om}{\omega} 
\newcommand{\pbh}{\mbox{PBH }}
\newcommand{\pbhs}{\mbox{PBHs }}
\newcommand{\mbh}{\mbox{$M_{\rm{BH}}$}}
\newcommand{\tbh}{\mbox{$T_{\rm{BH}}$}}
\newcommand{\trh}{\mbox{$T_{\rm{RH}}$}}

\newcommand{\mi}{\mbox{$M_{i}$}}

\newcommand{\th}{\mbox{$t_{\rm{H}}$}}
\newcommand{\mbhi}{\mbox{$M_{{\rm{BH}}i}$}}

\newcommand{\templ}{\mbox{$T_{\rm{Pl}}$}}

\newcommand{\mrh}{\mbox{$M_{\rm{RH}}$}}
\newcommand{\gm}{\mbox{$\gamma$}}

\newcommand{\obh}{\mbox{$\Omega_{\rm{BH}}$}}
\newcommand{\lqcd}{\mbox{$\Lambda_{\rm{QCD}}$}}

\newcommand{\s}{\mbox{$\rm{sec}$}} 
\newcommand{\g}{\mbox{$\rm{g}$}}
\newcommand{\mev}{\mbox{$\rm{MeV}$}}
\newcommand{\gev}{\mbox{$\rm{GeV}$}}
\newcommand{\mstar}{\mbox{$\rm{M_{\odot}}$}}

\newcommand{\deli}{\mbox{$\delta_{i}$}}
\newcommand{\simr}{\mbox{$\sigma_{\rm{R}}$}}
\newcommand{\simh}{\mbox{$\sigma_{\rm{H}}$}}

\newcommand{\mpl}{\mbox{$M_{\rm{Pl}}$}}
\newcommand{\mnr}{\mbox{Mon. Not. R. Astron. Soc.}}
\newcommand{\com}{\mbox{Commun. Math. Phys.}}
\newcommand{\lsim}{\lesssim}
\newcommand{\gsim}{\gtrsim}

\begin{document} 
\draft 


\title{Diffuse $\gamma$-ray background and
primordial black hole constraints on the spectral index of density fluctuations} 
\author{Hee Il Kim} 
\address
{The Research Institute for Natural Sciences, Hanyang University,
133-791, Seoul, Korea} 
\author{Chul H. Lee} 
\address 
{Department of
Physics, Hanyang University, 133-791, Seoul, Korea} 
\author{Jane H. MacGibbon}
\address{Code SN3, NASA Johnson Space Center, Houston, Texas 77058,
  USA}

\maketitle
\begin{abstract} 
We calculate the flux of $\gamma$-rays emitted from
primordial black holes (PBHs) which are formed by a ``blue'' power-law
spectrum of density fluctuations
in the early universe. Gamma-ray emission from such PBHs may
contribute significantly to the observed extragalactic diffuse $\gamma$-ray background
(DGB). Using the observed DGB flux from the imaging Compton
Telescope (COMPTEL) and the Energetic Gamma Ray Experiment Telescope
(EGRET) as the upper
limit of $\gamma$-ray flux from PBHs, we derive the upper limit on the
spectral index $n$ of the density fluctuations. The range of initial PBH masses which can contribute to the DGB is $2\times 10^{13}\g
- 5\times 10^{14}\g$, corresponding to a cosmic reheating temperature of $7\times 10^{7}\gev - 4\times 10^{8} \gev$. In this
range, we find the upper limit to be $n \lsim 1.23-1.25$. This limit
is stronger than those derived from the energy density in PBHs or PBH
relics and matches the value of $n$ required to explain the cosmic
microwave background anisotropy.
\end{abstract}

\pacs{95.85.Pw, 97.60.Lf, 98.80.-k}
\narrowtext


\newpage 
\section{Introduction} 
Distinct from black holes which form by recent processes like stellar collapses, black holes formed by mechanisms in
the early universe can exist. Such black holes, named primordial black holes
(PBHs), produce many interesting consequences in the early universe
and can also be sources of present astrophysical events. The simplest
mechanism for \pbh formation is the density fluctuations in the early
universe \cite{zel}. Overdense regions which strongly
deviate from the background universe can evolve into black holes
when the overdense regions enter the cosmological horizon. The resulting \pbh
mass is about the horizon mass at entry. Thus, \pbh masses can be
as small as about the Planck mass $\mpl \simeq 2\times 10^{-5}
\g$ or as large as $10^{7} \mstar $. In the latter case, they
can bound the mass of a typical galaxy after decoupling
\cite{rya}. \pbhs surviving today, or their massive relics, can be sources of dark matter.

Particle emission from black holes due to Hawking evaporation
\cite{haw} enlarges the role of PBHs. Interactions of the emitted
particles with the matter in the universe can affect numerous early
universe phenomena, such as nucleosynthesis \cite{miy}, baryogenesis
\cite{car76}, cosmic microwave background radiation (CMBR) distortion
\cite{nas}, entropy production \cite{zelsta}, diffuse $\gamma$-ray
background (DGB) \cite{car76,paghaw,maccar,hal}, and so on. \pbhs
with initial mass $\sim 5\times 10^{14} \g$ are presently at the final stage of their
evaporation \cite{mac91} and may emit enormous amounts of energy. In some
GUT-scale theories, the relics of \pbhs created with initial mass $\lsim 5\times
10^{14} \g$ and expired by today can constitute the dark matter \cite{mac87}.

From the effects of \pbhs mentioned above, upper limits on $\obh$, the
fraction of the critical energy density of the universe which can be in PBHs, are found
\cite{nov}. Constraints on the spectral index of the density
fluctuations have been derived from these energy density limits
\cite{cargil,mp2} and it has been shown that \pbhs in general give the
strongest upper bounds on the spectral index \cite{gre}. 

The upper
limit of $\obh$ in PBHs with masses about $5\times 10^{14} \g$ was
also calculated from the PBH 
$\gamma$-ray emission \cite{paghaw,maccar}. There exists a
homogeneous and isotropic $\gamma$-ray background in the universe
whose origin is known to be extragalactic \cite{ram,fic}. Earlier
authors postulated a contribution to the DGB from the PBH 
$\gamma$-ray emission \cite{car76,paghaw,maccar}.  MacGibbon and Carr have recently
updated the limit on the PBH density using the DGB measurement of the Energetic Gamma Ray Experiment Telescope
(EGRET) \cite{sre} and found that $\obh \lsim (5.1\pm1.3)\times
10^{-9} h^{-1.95\pm 0.15}$ \cite{pre}. Here $h$ is the Hubble parameter in
units of $100 \rm{km s^{-1} Mpc^{-1}}$. Similar values are deduced
from the PBH antiproton and $\gamma$-ray emission if PBHs
cluster along with cold dark matter in galactic halos \cite{mak,wri}.
The limits on $\obh$ in this mass range assume, though, 
that the density fluctuations have a scale-invariant Harrison-Zel'dovich
spectrum with spectral index $n=1$. Therefore they
cannot be converted into an upper limit on the spectral index. Also,
in these approaches the fluctuation amplitude was not explicitly normalized. Instead,
a parameter related to the PBH density was introduced and varied to
match the PBH $\gamma$-ray flux to the observed DGB. However, if one 
normalizes the fluctuation amplitude on the scales for PBH
formation to that detected by COBE on much larger scales from the CMBR anisotropy
amplitude $\delta \sim 1.9\times 10^{-5}$ \cite{gor}, it is impossible to form a significant number of \pbhs
with a continuous Harrison-Zel'dovich spectrum. With the normalization to the CMBR
anisotropy, significant \pbh abundance is possible only if the density
fluctuations have an $n >1$ (``blue'') spectrum. A blue spectrum with a
constant spectral index is a valid assumption, for example, in the
hybrid inflationary scenario \cite{cargil}. 

For these reasons, we will
re-examine the $\gamma$-ray flux from \pbhs formed by an $n > 1$
spectrum and find the upper limit on the spectral index, using the
recent DGB measurements of the imaging Compton Telescope (COMPTEL)
\cite{kap} and the EGRET
\cite{sre} on board the Compton Gamma Ray Observatory (CGRO) and normalizing the fluctuation amplitude to that on CMBR anisotropy scales. While
previous authors \cite{cargil,gre} have estimated limits on $n$ from the
$\gamma$-ray emission of PBHs formed by a blue fluctuation spectrum,
they have not performed the explicit calculation of the PBH
$\gamma$-ray emission and matched it to the DGB, analogous to the
approach of MacGibbon and Carr for $n=1$ \cite{maccar,pre}. They also did
not include the effect of quark and gluon production which dominates
the emission above black hole temperatures of about $100 \mev$
\cite{macweb}. 

In
Sec. II, we review the \pbh formation and PBH mass spectrum,
correcting errata in Ref. \cite{mp2}. The black hole evaporation
and QCD fragmentation effects on the particle emission are discussed in
Sec. III and the calculation of the $\gamma$-ray flux from \pbhs is
given in Sec. IV. The recent DGB observations are reviewed in
Sec. V. Our detailed calculation of the upper limit on $n$ is
presented in Sec. VI. The paper closes with
some concluding remarks in Sec. VII.

\section{PBHs and their mass spectrum} 
We address PBH
formation in a universe with a hard equation of state, that is $p=\gamma
\rho$ with $0 < \gamma \lsim 1$. Studies of the
evolution of a spherical overdense region whose initial radius $R$
is greater than the particle horizon show
that for the region to collapse to a black hole, the
initial density contrast of the region, $\deli$, should satisfy the
following condition \cite{car75} 
\be \beta^{2}
\biggr(\frac{\mi}{M_{{\rm{H}}i}}\biggr)^{-\frac{2}{3}} \lsim \delta_{i}
\lsim \alpha^{2} \biggr(\frac{\mi}{M_{{\rm{H}}i}}\biggr)^{-\frac{2}{3}}~~~.
\ee 
Here $\alpha$ and $\beta$ are constants of the order of
$\sqrt{\gamma}$, $\mi$ is the mass contained in the region of radius
$R$ at the time $t_{i}$ when the fluctuation develops, and $M_{{\rm{H}}i}$
is the horizon mass at $t_{i}$. The lower bound comes from the
requirement that the radius of the region at its maximum expansion
should be larger than the Jean's length at that epoch. The upper bound
comes from the requirement that the overdense region should not be disconnected from the universe.
Since $\mi \propto R^{3}$ and $R \propto k^{-1}$, we will use $\mi, R$ and
$k$ interchangeably to represent the initial mass, size or comoving
wavelength.  The \pbhs form when the
overdense region enters into the horizon. The resulting \pbh mass,
$\mbhi$, is approximately the horizon mass at that time $\th$ and is given by 
\be
\mbhi \simeq \gamma^{3/2} M_{{\rm{H}}i}\frac{\th}{t_{i}}~~~.  
\ee 
$\mbhi$ is related to $\mi$ via \cite{car75}
\be 
\mbhi \simeq
\gamma^{3\gamma/(1+3\gamma)}M^{(1+\gamma)/(1+3\gamma)}_{i}M^{2\gamma/(1+3\gamma)}_{{\rm{H}}i}~~~.
\ee 
As the universe expands, larger \pbhs are
formed, so that \pbhs with masses less than $\mbhi$ coexist in the universe at time $\th$.

We assume the density fluctuations to be Gaussian. Recently it was claimed that a non-Gaussian nature to the fluctuations
may affect the \pbh formation \cite{bul}. This is model dependent and does not much alter
the upper limit on $n$ \cite{gre}. If one surveys the universe with a window having size $R$, the smoothed density field $\delta_{R}({\bf{x}})$ is defined
by 
\be 
\delta_{R}({\bf{x}})=\int d^{3}{\bf{y}}
\delta({\bf{x}}+{\bf{y}})W_{R}({\bf{y}})~~~,  
\ee 
where
$\delta({\bf{x}}) \equiv (\rho({\bf{x}})-\rho_{b})/\rho_{b}$, $\rho_{b}$
is the background energy density of the universe, and $W_{R}({\bf{x}})$
is the smoothing window function of size $R$. The dispersion $\simr$, the standard deviation of the density contrast of the regions with
$R$, is given by 
\be \sigma_{R}^{2}=\frac{1}{V^{2}_{W}}\langle
\delta_{R}^{2}({\bf{x}}) \rangle = \frac{1}{V^{2}_{W}}\int
\frac{d^{3}{\bf{k}}}{(2\pi)^{3}}|\delta_{{\bf{k}}}|^{2}W_{{\bf{k}}}^{2}(R)~~~.
\ee 
where $V_{W} \sim R^{3}$ denotes the effective volume filtered by
$W_{R}$, and $\delta_{{\bf{k}}}$ and $W_{{\bf{k}}}$ are the Fourier
transforms of $\delta(\bf{x})$ and $W_{R}(\bf{x})$, respectively.
For Gaussian fluctuations the probability that the region of size $R$ has density
contrast in the range $(\delta+d\delta, \delta)$ is 
\be P(\mi,
\delta)d\delta=\frac{1}{\sqrt{2\pi}\simr}\exp
\biggr(-\frac{\delta^{2}} {2\sigma^{2}_{R}}\biggr) d\delta~~~.  
\ee

Thus, the probability that the region with $\mi$ collapses to a black
hole is
\be P_{\rm{BH}}(\mi)=
\int^{{\cal{A}}}_{{\cal{B}}}P(\mi, \delta)d\delta~~~, 
\ee 
with
\be
{\cal{A}}=\alpha^{2}
\biggr(\frac{\mi}{M_{{\rm{H}}i}}\biggr)^{-\frac{2}{3}},~~~~~
{\cal{B}}=\beta^{2}
\biggr(\frac{\mi}{M_{{\rm{H}}i}}\biggr)^{-\frac{2}{3}}~~~.  
\ee 
The above quantity $P_{\rm{BH}}(M)$ has been interpreted as the ratio
of the density in PBHs to the density of the universe. This is not
strictly so
because regions larger than $\mi$ also contribute. As in
our previous work \cite{mp2}, we proceed by omitting these
contributions and find the number density of black holes produced by
the collapse of regions with mass between $\mi$ and $\mi+d\mi$ to be \cite{mp2}
\be
n_{\rm{BH}}(\mi)d\mi=-\frac{\rho_{i}}{\mi}\sqrt{\frac{2}{\pi}}
\frac{\cal{B}}{\sigma_{R}^{2}}\frac{\partial \sigma_{R}}{\partial \mi}
\exp \biggr(-\frac{{\cal{B}}^{2}}{2\sigma_{R}^{2}}\biggr) d\mi~~~,  
\ee
Here $\rho_{i}=3/(32\pi G t^{2}_{i})$ is the background energy density of the universe at $t_{i}$.
Unlike Ref. \cite{mp2}, we will not convert the mass spectrum into
a function of unsmoothed quantities but retain the use of the
smoothed quantities. This is to avoid factors being dropped in the
conversion.

Since fluctuations grow as $a^{2}(t)$ in the
radiation-dominated era, where $a(t)$ denotes the cosmic expansion factor,
the dispersion corresponding to $\simr$ becomes at horizon crossing
\be
\sigma_{\rm{H}}=\biggr(\frac{\mi}{M_{{\rm{H}}i}}\biggr)^{\frac{2}{3}}\simr~~~,
\ee 
whence $n_{\rm{BH}}(\mi)$ can be written as 
\be
n_{\rm{BH}}(\mi)d\mi=-\sqrt{\frac{2}{\pi}}\frac{\rho_{i}}{\mi}\gamma
\biggr[\frac{1}{\sigma^{2}_{\rm{H}}}\frac{\partial \simh}{\partial
\mi}-\frac{2}{3}\frac{\sigma^{-1}_{\rm{H}}}{\mi}\biggr] \exp
\biggr(-\frac{\gamma^{2}}{2\sigma^{2}_{\rm{H}}}\biggr)d\mi~~~.  
\ee
Applying Eq. (3) with $\gamma=1/3$, the mass spectrum can be expressed as a function
of the initial \pbh mass $\mbhi$ and we have
\be
n_{\rm{BH}}(\mbhi)d\mbhi=-\sqrt{\frac{2}{\pi}}\gamma^{\frac{7}{4}}\rho_{i}
M^{\frac{1}{2}}_{\rm{H}i}M^{-\frac{3}{2}}_{\rm{BH}i}
\biggr[\frac{1}{\sigma^{2}_{\rm{H}}}\frac{\partial \simh}{\partial
\mbhi}-\frac{\sigma^{-1}_{\rm{H}}}{\mbhi}\biggr] \exp
\biggr(-\frac{\gamma^{2}}{2\sigma^{2}_{\rm{H}}}\biggr)d\mbhi~~~.  
\ee

In general, tensor perturbations (gravitational waves) are also
produced in inflationary scenarios and contribute to the CMBR
anisotropy. However, inclusion of the tensor perturbations does not
significantly affect the PBH mass spectrum if the fraction in tensor
perturbations is not dominant \cite{cargil}. Therefore we assume that the
anisotropy is only due to the scalar fluctuation. 

The four-year results of the COBE experiment resolve the horizon crossing
amplitude at present to be \cite{gor} 
\be \delta_{0}=1.91\times
10^{-5}\exp(1.01(1-n))~~~, 
\ee 
with $n=1.2\pm 0.3$.  This implies a smoothed
amplitude today $\sigma_{0}$ of $9.5\times 10^{-5}$
with a slight dependence on $n$ \cite{gre}. We denote 
the mass contained at $t_{i}$ in the region whose comoving scale corresponds to the present horizon scale by $M_{0}$.
Under the power-law spectrum assumption, $\simh \propto k^{(n-1)/2}$
and so 
\be
\simh=\sigma_{0}\biggr(\frac{\mi}{M_{0}}\biggr)^{\frac{1-n}{6}} 
\ee
(Note that the
spectral index $n$ of Ref. \cite{car75} is equivalent to $(n+3)/6$ in this
work.)
$M_{0}$ is not the present horizon mass as it was incorrectly
taken to be in Ref. \cite{mp2}. That misidentification introduced further errors
when converting $\simh$ into a function of $\mbhi$ and made the results
of Ref. \cite{mp2} far weaker than those of Ref. \cite{gre}. Our revised
results are given in Sec. VI. 

From Eq. (3), $\simh$ can be represented
as a function of $\mbhi$ 
\be
\simh=\sigma_{0}\biggr(\frac{\mbhi}{\mbh_{i0}}\biggr)^{p}~~~,
\ee 
with $p=(1-n)/4$ in the radiation-dominated era ($\gm=1/3$) and
$p=(1-n)/6$ in the matter-dominated era ($\gm=0$). Minor
discrepancies, which arise in Eq. (15) at the transition into the
matter-dominated era, lead to less than a $1\%$ change in the
constraint on $n$ \cite{gre}.
 
The PBH initial mass spectrum under the $n>1$ power-law
spectrum assumption is thus described by 
\be n_{\rm{BH}}(\mbhi)d\mbhi=
\frac{n+3}{4}\sqrt{\frac{2}{\pi}}\gamma^{\frac{7}{4}}\rho_{i}
M^{\frac{1}{2}}_{{\rm{H}}i}M^{-\frac{5}{2}}_{{\rm{BH}}i}
\sigma^{-1}_{\rm{H}} \exp
\biggr(-\frac{\gamma^{2}}{2\sigma^{2}_{\rm{H}}}\biggr)d\mbhi~~~.  
\ee

\section{particle emission from black holes} 
Due to the Hawking effect \cite{haw}, a
rotating charged black hole emits particles at a rate, 
\be
\frac{dN_{s}}{d\om
dt}=\frac{\Gamma_{s}}{2\pi}\biggr[\exp\biggr(\frac{\om-l\Omega-q\Phi}
{\kappa/2\pi}\biggr)+(-1)^{2s}\biggr]^{-1}~~~, 
\ee 
per degree of particle
freedom. Here, $\kappa, \Omega$, and $\Phi$ are the surface gravity,
angular velocity and electric potential, respectively, $s$ is the
particle spin, $l$ is the axial quantum number or angular momentum and
$q$ is the particle charge. The absorption probability for the emitted
species, $\Gamma_{s}$, is in general a function of $\om, \Omega, \Phi,
\kappa$, together with the internal degrees of freedom and rest mass
of the emitted particle. Taking the above emission rate, it has been shown that
$\Omega \rightarrow 0$ before most of the black hole evaporates \cite{pag76} and
that a black hole with mass $\lsim 10^{6}\mstar$ discharges faster
than it evaporates \cite{gib}. Hence, it is natural to regard \pbhs as Schwarzschild black holes.

At high energies, $\Gamma_{s}
\propto \om^{2}$ for massless or relativistic particles and the
emission spectrum mimics the radiation from a black body of temperature
$\tbh= \kappa/2\pi$. Noting that the surface gravity is
$\kappa=1/4G\mbh$ when $\Omega=\Phi=0$, the temperature of a black hole can be
defined as 
\be 
\tbh=\frac{1}{8\pi G\mbh} \simeq
1.06\biggr(\frac{\mbh}{10^{13}\g}\biggr)^{-1} \gev~~~. 
\ee 
At low energies,
$\Gamma_{s}$ does not simply scale as $\om^{2}$ but depends on other
quantities mentioned above. The form of
$\Gamma_{s}$ has been explored both analytically and numerically
(see references in \cite{bir}). Extensive numerical studies
of the direct particle emission from black holes were done by Page
\cite{pagpag}. Hawking emission can be thought of as a
process by which a black hole emits particles with approximately a blackbody
radiation spectrum once the black hole temperature exceeds the
rest mass of the particle. Thus, black holes with masses larger than
$10^{17}\g$, corresponding to $\tbh \simeq 0.1 \mev$, emit only massless
particles. As the black hole mass decreases, massive particles
will be emitted.

In the conventional viewpoint, it is natural to assume that
elementary particles like quarks and gluons, rather than composite hadrons, are
directly emitted from black holes once the emission energy exceeds the QCD confinement scale, $\lqcd$. In this picture, pions are only directly emitted
from black holes in the energy range between $100\mev$ and 
$\lqcd$ and are produced by quark and gluon decay above $\lqcd$. Taking into account the number of the
emitted species, the mass loss rate of a black hole can be written as \cite{mac91} 
\be
\frac{d\mbh}{dt}=-5.34\times 10^{25}\phi(\mbh)M^{-2}_{\rm{BH}} \g~
\s^{-1}~~~, 
\ee 
where $\phi(\mbh)$, a function of the number of directly emitted
species, is normalized to unity for $\mbh \gg 10^{17} \g $. Black
holes with masses $5\times 10^{14}\g \ll \mbh \ll 10^{17}\g$ emit
$e^{\pm}$, neutrinos and photons and have initially $\phi(\mbh)=1.569$. If
black holes can emit three lepton families, six quark flavors, the
photon and direct pions, then $\phi(\mbh)\lsim 13.9$. Including the
emission of weak gauge and higgs bosons, $\phi(\mbh)\lsim 15.4$ for
$\tbh \lsim 100\gev$ and $\mbh
\gsim 10^{11}\g$. At higher energies or in non-standard models
like supersymmetry or superstrings, $\phi(\mbh)$ may be greater but in
general remains less than 100. 

Integrating Eq.(19),
the black hole lifetime, $\tau_{\rm{evap}}$, is found to be \cite{mac91} 
\ba
\tau_{\rm{evap}}&\simeq& 1.2\times
10^{3}\frac{G^{2}M^{3}_{\rm{BH}}}{\phi(\mbh)} \no
           &=&6.24\times 10^{-27}M^{3}_{\rm{BH}}\phi(\mbh)^{-1} \s~~~.
\ea

The jet-like fragmentation and hadronization of the quarks and gluons evaporated above
$\lqcd$ drastically change the observable spectrum of emitted particles. The
evaporated quarks and gluons fragment into further quarks and gluons which then
compose themselves into hadrons on distances greater than
$\Lambda^{-1}_{\rm{QCD}}$ in the jet frame. These particles further decay into
the astrophysically stable particles - photons, neutrinos, electrons and
protons and their antiparticles. MacGibbon and Webber have shown that this picture is analogous
to the decay of quark and gluon jets in $e^+e^-$ accelerator events and
calculate the instantaneous flux of particles for $0.2\gev \lsim \tbh \lsim 100
\gev$ by convolving Eq. (17)
with the HERWIG QCD jet code \cite{macweb}. Their results differ strongly from those of
previous works which omitted QCD emissions and particle
decays. They find that the black hole emission at these temperatures is dominated by the jet fragmentation
products. In the case of photons, the primary peak in the black hole emission is due to the decay of
jet-produced $\pi^{0}$ and occurs around $67 \mev$. The position of the photon peak
does not shift significantly with the black hole temperature. In
contrast, the photons directly emitted by the black hole, not resulting from jet decay, appear at $\simeq 5\tbh$ with fluxes 4 to 5 orders less than
the flux at the jet-dominated peak. The jet-produced photons were
omitted in previous estimates of the \gm-ray emission from PBHs formed
by $n>1$ density fluctuations \cite{cargil,gre}. In this paper, we fit
the instantaneous $\gamma$-ray fluxes from $\tbh=0.2-100 \gev$ black holes which are shown
in Fig. 4 of Ref. \cite{macweb} and derived including quark and gluon
emission. We then use these results to calculate the $\gamma$-ray flux from PBHs.

\section{Gamma-rays from PBHs} 
To calculate the flux of $\gamma$-rays
from PBHs, it is important to know how many \pbhs have existed in the
universe. This can be determined from the PBH initial mass spectrum,
which in turn
depends on the fluctuation amplitude and spectral index, and
$t_{i}$, the time when the fluctuations develop. Previous works
 considered only the case in which the fluctuations are described by the
Harrison-Zel'dovich spectrum and hid the effect of fluctuation amplitude
and $t_{i}$ \cite{car76,paghaw,maccar,hal}. 
However, it is impossible to form a significant number of \pbhs
with the Harrison-Zel'dovich spectrum if the fluctuation
amplitude on PBH formation scales is normalized to the amplitude found by
the COBE experiment on much larger scales. With normalization to the CMBR anisotropy, substantial \pbh formation is possible only if the fluctuations increase on small
scales. This occurs if the fluctuations satisfy an $n > 1$ power-law spectrum. We will assume that
the fluctuations follow a power-law spectrum and calculate the $\gamma$-ray
flux from \pbhs with the initial mass spectrum given by Eq. (16).

Since the number density of \pbhs decreases as $R^{-3}$, the number
density of \pbhs at the time $t_{1}\geq \th$ is 
\be
n^{\prime}_{\rm{BH}}(t_{1}) \simeq \biggr(\frac{R_{1}}{R_{i}}\biggr)^{-3}\int^{M_{{\rm{BH}}1}}_{M_{\ast}(t_{1})}n_{\rm{BH}}(\mbhi)d\mbhi~~~,
\ee 
where $M_{\ast}(t_{1})$, the initial mass of a PBH whose lifetime is $t_{1}$, is given by 
\be 
M_{\ast}(t_{1}) \simeq
\biggr[\frac{\phi(M_{\ast}(t_{1}))}{6.24\times
10^{-27}}\biggr(\frac{t_{1}}{1\s}\biggr)\biggr]^{1/3} \g~~~.
\ee 
We denote
the instantaneous $\gamma$-ray flux from a black hole with mass $\mbh$ as
$f_{\gamma}(\mbh,\om)$. At $t_{1}$, \pbhs with initial mass $\mbhi$
have evaporated down to a mass
\be 
M_{\rm{evap}}\simeq (M^{3}_{{\rm{BH}}i}-1.6\times
10^{26}\phi(\mbhi)t_{1})^{1/3} 
\ee 
and are emitting photons with
flux $f_{\gamma}(M_{\rm{evap}},\om)$. The angular frequency at
emission, $\omega$, is redshifted by the expansion of the universe to
a present
angular frequency $\omega_{0}$ of 
\be
\omega=\frac{R_{0}}{R_{1}}\omega_{0}~~~.
\ee 
Thus the total flux per unit solid angle at today, $t_{0}$, of
\gm-rays emitted from PBHs is
\be
\frac{dJ}{d\omega_{0}}=\frac{1}{4\pi}\int^{t_{0}}_{t_{min}}\biggr(\frac{R_{0}}{R_{1}}\biggr)
\biggr(\frac{R_{1}}{R_{i}}\biggr)^{-3}dt_{1}
\int^{M_{{\rm{BH}}1}}_{M_{\ast}(t_{1})}f_{\gamma}(M_{\rm{evap}},\om)n_{\rm{BH}}(\mbhi)d\mbhi~~~
\ee 
where $t_{min}$ is the earliest time after inflation at which PBHs
form.

Photons emitted by the black holes may
interact with ambient matter in the universe via many processes and
lose energy or be cut off during propagation. If a photon effectively interacts
$\tau$ times during flight, the photon flux is
attenuated by a factor of $e^{-\tau}$. Therefore, the actual photon
flux reaching Earth at the present time is 
\be
\frac{dJ}{d\omega_{0}}=\frac{1}{4\pi}\int^{t_{0}}_{t_{min}}\biggr(\frac{R_{0}}{R_{1}}\biggr)
\biggr(\frac{R_{1}}{R_{i}}\biggr)^{-3}dt_{1}
\int^{M_{{\rm{BH}}1}}_{M_{\ast}(t_{1})}
e^{-\tau}f_{\gamma}(M_{\rm{evap}},\om)n_{\rm{BH}}(\mbhi)d\mbhi~~~.
\ee
The number of interactions $\tau$, known as the optical depth,
depends on the energy and the time or redshift $z$ at emission. A detailed treatment on
the optical depth is given in Sec. VI.

\section{Extragalactic diffuse gamma-ray background} 
Since its first
discovery in the $0.1-2\mev$ range by detectors on the lunar probes \cite{arn}, the homogeneous and isotropic diffuse $\gamma$-ray
flux, whose origin is extragalactic, has been observed in numerous
satellite and balloon-borne experiments.
The SAS-2 satellite provided the first clear evidence for the
existence of an extragalactic $\gamma$-ray background between $30-150
\mev$ \cite{fic}. Several Apollo and balloon-borne experiments also saw evidence of a bump in the few $\mev$ range in excess of
the extrapolated X-ray continuum \cite{tro}.

A number of models were proposed to explain
the early measurements of the extragalactic spectrum. Active galaxies,
which can be observable sources of
discrete extragalactic $\gamma$-ray emission when they are located close to
our Galaxy, are believed to contribute at least in part
\cite{str}. Another model which has been considered is
matter-antimatter annihilation at the boundaries of superclusters
\cite{ste}. In this model, the MeV bump is attributed to the redshifted
 peak of $\pi^{0}$ decays at $67 \mev$. PBH $\gamma$-ray emission has
 also been proposed as a contributor to the DGB flux. In some exotic
 models, PBH emission may additionally explain the MeV bump
 \cite{maccar}. However, none of the scenarios for the DGB production,
 by themselves, is sufficient to
explain the measured flux and spectrum.

With higher sensitivity and wide-field of view, the detectors on the
CGRO enlarge the detection range and gather important
data on the DGB. The DGB flux measured
by the COMPTEL \cite{kap} at $1 - 30 \mev$ is now compatible with power-law extrapolations
of the measured flux at lower and higher energies. The results below about $9 \mev$
are preliminary but the $2-9 \mev$ flux is far less
than previously measured and no $\mev$ bump is seen in
this region at the levels reported previously. This weakens the need
to explain an MeV feature. 

We parameterize the preliminary COMPTEL results \cite{kap} in the range
$0.8-30 \mev$ by the best-fit power law function given in Ref. \cite{kri} 
\be
\frac{dJ}{d\om_{0}}\biggr|_{\rm{obs}}= 6.40\times
10^{-3}\biggr(\frac{\om_{0}}{1\mev}\biggr)^{-2.38}~~~{\rm{[cm^{2}~s~sr~MeV]^{-1}}}~~(\rm{COMPTEL})~~~.
\ee
In the
range $30\mev - 100 \gev$, the EGRET experiment finds the DGB flux to be well described by the single
power-law function \cite{sre}, 
\be
\frac{dJ}{d\om_{0}}\biggr|_{\rm{obs}}=
k\biggr(\frac{\omega_{0}}{451\mev}\biggr)^{-\alpha}~~~~~ {\rm{[cm^{2}~s~sr~MeV]^{-1}}}~~~(\rm{EGRET})
\ee 
with
$k=(7.32\pm0.34)\times 10^{-9}$ and $\alpha=2.10\pm 0.03$. No large
scale spatial anisotropy or deviations in the energy spectrum is
discernible in the extragalactic component above $30\mev$. The observed flux above
$10 \mev$, and possibly up to $100\gev$, may be explained by unresolved 
blazers \cite{sre}. Below $10 \mev$, the measurements still have large
uncertainties and the exact nature of the emission is
not well understood.

\section{Constraints on the spectral index}

We now derive the constraints on the spectral index
from the condition that the PBH $\gamma$-ray flux should not be
larger than the observed DGB flux. With the normalization of the
fluctuation amplitude to the CMBR anisotropy, \pbh
formation is limited to the epoch when the fluctuation arises. This time is related to the reheating
temperature in inflationary models by \cite{kol} 
\be t_{i{\rm{RH}}}=0.301
g^{-1/2}_{*}\frac{\mpl}{T^{2}_{\rm{RH}}}\sim
\biggr(\frac{T_{\rm{RH}}}{\mev}\biggr)^{-2} \s~~~.
\ee 
Here $g_{*}\sim
100$ counts the degrees of freedom of the constituents in the early
universe. The minimum initial \pbh mass corresponding to $T_{\rm{RH}}$ is 
\be
\mrh \simeq \frac{1}{8}\gamma^{3/2}\mpl \biggr(\frac{\trh}{\templ}\biggr)^{-2}~~~.
\ee 
PBHs created before the onset of reheating will be diluted to an
insignificant density during inflation.
Since we are considering the case $n>1$, the
resulting PBH initial mass spectrum has a very narrow mass range. Thus
we will make the approximation that the photons are solely emitted by
PBHs whose initial mass is $\mrh$.

The photon interactions in the matter-dominated era,
relevant to the DGB observations, are Compton
scattering, pair production, photo-ionization, and photon-photon interactions. Via
these processes and cosmological redshift, the energy of the emitted
photons is degraded. Zdziarski and Svensson have studied the attenuation of
$\gamma$-ray flux at cosmological distance \cite{zdz}. They found that
the maximum redshift from which photons can be detected today peaks at
$z_{max}\simeq 700$ and for present energies $1\mev \lsim \omega_{0} \lsim 1 \gev$
. All photons emitted at higher redshifts are cut off by
interactions (i.e, $\tau(\omega_{0},z)>1$) and do not reach
Earth. This means that \pbhs which completely evaporated before
$z_{max} \simeq 700$
can not contribute to the present DGB. From Eq. (20), this corresponds to a
minimum detectable initial PBH mass of about $2\times 10^{13} \g$ and a
reheating temperature of
$\trh \simeq 4\times 10^{8}\gev$. Noting that $M_{\ast}(t_{0}) \simeq
5\times 10^{14} \g$, the range of \pbhs which can contribute to the
observed DGB is then $2\times
10^{13}\g - 5\times 10^{14}\g$ and the corresponding reheating
temperature range is $7\times 10^{7}\gev - 4\times 10^{8}
\gev$. Outside the range $1\mev \lsim \omega_{0} \lsim 1 \gev$, the
maximum redshift for a given energy, $z_{max}(\omega_{0})$, is less
than 700 and depends on the details of the interactions.  We take
$z_{max}(\omega_{0})$ from Ref. \cite{zdz}. Only photons for which
$\tau<1$ are here included and we regard these photons as being free
from attenuation. The integrated $\gamma$-ray flux from PBHs does not
depend significantly, though, on the details of the optical depth.

We now proceed to calculate the integrated PBH $\gamma$-ray flux, Eq. (26), using the
instantaneous emission from individual black holes, $f_{\gamma}$, obtained by fitting the simulations of
Ref. \cite{macweb}. 
Our
results are shown in Fig. 1. It can readily be seen that the
$\gamma$-ray flux from PBHs can not fully explain the observed DGB
flux although the PBH emission may contribute significantly to the observed DGB flux around
$10-100\mev$. In the $n>1$ case, the PBH flux arises from the lifetime
emission of PBHs with initial mass $\mrh$, whereas the
Harrison-Zel'dovich spectrum produces a broad range of initial
PBH masses. In both cases, the PBH flux falls off as roughly $\omega^{-3}_{0}$
above $100\mev$
\cite{car76,paghaw,maccar}. This high energy tail mainly comes from the lifetime
direct photon emission in the most recent evaporation epoch
\cite{mac91}. At low energies, where the flux is strongly determined by
QCD jet fragmentation, the flux spectrum for $n>1$
does not scale as $\omega^{-1}_{0}$, as for $n=1$,
but instead flattens out due to the narrow
PBH initial mass range. In addition, the turnover in the spectrum
occurs at lower energies as the reheating temperature increases and
$\mrh$ decreases. This is because the turnover corresponds to the
redshifted peak emission of an $\mrh$ black hole emitting at its
initial temperature. 

From the
constraint that the PBH $\gamma$-ray flux can not
be larger than the observed DGB flux, we derive the upper limits of the
spectral index in the range $7\times 10^{7}\gev \lsim \trh \lsim
4\times 10^{8}\gev$ (Fig. 2). The upper limit on $n$ is 
\be
n \lsim 1.23-1.25~~~.
\ee
Even though photons emitted in models with higher $\trh$ suffer more
interactions and larger redshifts, the exponential dependence of the
PBH mass spectrum, Eq. (16), implies that the number density of PBHs
strongly increases with $\trh$ for a given spectral index. Thus, the
upper limit on $n$ decreases as the reheating temperature grows. That
is, the constraint on $n$ becomes stronger as the reheating
temperature increases. These values are similar to the upper limits
obtained from the deuterium destruction constraint in the lower mass
range  $10^{9}\g \lsim \mbh \lsim 10^{13}\g$ \cite{gre}. 

Also shown in Fig. 2 are the weaker limits on $n$ derived from the
maximum allowable energy density in PBHs. We plot the upper limits found from the requirement
that the PBH energy density does not overclose the universe at any
epoch, $\obh<1$ (Case I, the dashed line), and the similar restriction
on any present relic density in PBHs which did not evaporate
completely but left residual masses of about the Planck
mass, $\Omega_{\rm{relic}}<1$ (Case II, the dotted line). The latter
constraint strengthens somewhat if the relic mass is greater than
$\mpl$ \cite{cargil,mp2}. Constraint Case I applies regardless of whether PBHs evolve into massive relics. The
new upper limits on $n$ from the energy densities are much tighter
than those of Ref. \cite{cargil,mp2} and decrease as $\trh$ grows. In
Ref. \cite{gre}, upper limits on $n$ were found from the condition
that the present PBH density and any relic density satisfy
$\obh_{0}<1$ and $\Omega_{\rm{relic}}<1$, respectively. There it was
shown that the constraint $\obh_{0}<1$ is weaker than the relic
constraint, Case II. However, when we now extend that condition to
require that the PBH density fraction at any epoch does not overclose
the universe, we find that Case I is stronger than Case II if
$\trh \lsim 10^{13}\gev$. The
energy density in PBHs or PBH relics give weaker constraints than the DGB
because the upper limit on $\obh$ from the PBH $\gamma$-ray flux is
far less than 1. Because of the vast difference between the scales on
which the CMBR anisotropy occurs and the scales on which PBHs form, the
limit on the spectral index obtained from PBH emission is highly
insensitive to the true value of the CMBR anisotropy or $\obh$.

\section{Conclusions} 
In this paper, we calculate the $\gamma$-ray flux from
PBHs formed by density fluctuations in the early universe and compare
it with the observed extragalactic DGB flux. Previous works
considered the case in which the density fluctuations have an $n=1$
Harrison-Zel'dovich spectrum and did not explicitly normalize the fluctuation amplitude. If the fluctuation
amplitude on the scales of PBH formation is normalized to that on
large scales deduced from the COBE observations of the CMBR anisotropy,
PBHs can not form in cosmologically significant numbers from
a Harrison-Zel'dovich spectrum. Thus, we describe the fluctuations by an $n>1$ power-law
spectrum and find the upper limit on $n$ from the condition that the
$\gamma$-ray flux from PBHs should not be larger than the DGB flux. The smallness of the fluctuation amplitude from the
CMBR anisotropy limits PBH formation to
the time when the fluctuations develop. In inflationary models, this
time is related to the reheating time.  To
 model the $\gamma$-ray emission, we fit previous simulations of the
emission from individual black holes which included QCD fragmentation
and decays. Due to the
interactions of $\gamma$-rays with the background matter in the
universe, only PBHs surviving later than $z_{max} \lsim 700$ can contribute
to the DGB flux observed today. The initial mass range of PBHs
relevant to the observed 
DGB is then $2\times 10^{13}\g \lsim \mbh
\lsim 5\times 10^{14}$ with a corresponding cosmic reheating
temperature between $7\times 10^{7}\gev \lsim \trh \lsim
4\times 10^{8}\gev$. We find the resulting upper limit
on $n$ in this range to be $n \lsim 1.23-1.25$. Our constraint on $n$
is stronger than those obtained by requiring that the energy
density in PBHs does not overclose the universe at any epoch ($\obh
<1$) and that found by requiring that any present PBH relic density similarly does not
overclose the universe ($\Omega_{\rm{relic}} <1$). The upper limit
on $\obh$ implied by the PBH $\gamma$-ray flux is far less than 1. If the fluctuation
amplitude is constrained by the CMBR anisotropy and PBH emission, the
upper limit on $n$ is fine-tuned and highly insensitive to the
precise upper limit on $\obh$ or the precision in the CMBR
measurement.

Recently, Niemeyer and Jedamzik \cite{nie} have argued, supported by
preliminary numerical simulations, that sub-horizon mass PBHs may form
in considerable numbers at any formation epoch. For Gaussian
fluctuations, they deduce that the PBH initial mass distribution at a 
given formation time peaks at about 0.6 times the horizon mass and
extends from much smaller masses up to the horizon mass. Such a
distribution would have an effect on our limits similar to raising $\trh$.

We also note that it has been proposed that the emitted quarks and gluons from
a black hole may interact and form a photosphere around the black hole
above black hole temperatures of a few GeV
\cite{hec1}. This scenario, however, remains controversial. In the photosphere model, the flux would be more concentrated around
$100\mev$ than if the quarks and gluons directly
fragment into hadrons \cite{hec2}. The high energy tail from the PBH
distribution would also scale as $\om^{-4}_{0}$, not $\om^{-3}_{0}$.
Photosphere formation may somewhat weaken the constraint
on $n$ but such changes would be small due to the fine-tuned nature
of the constraint.

\acknowledgments
This work was supported in part by the Basic Science Institute Program, Korea Ministry of Education (Project No. BSRI-97-2441), and in part by the Korea Science and Engineering Foundation through the SRC program of SNU-CTP.

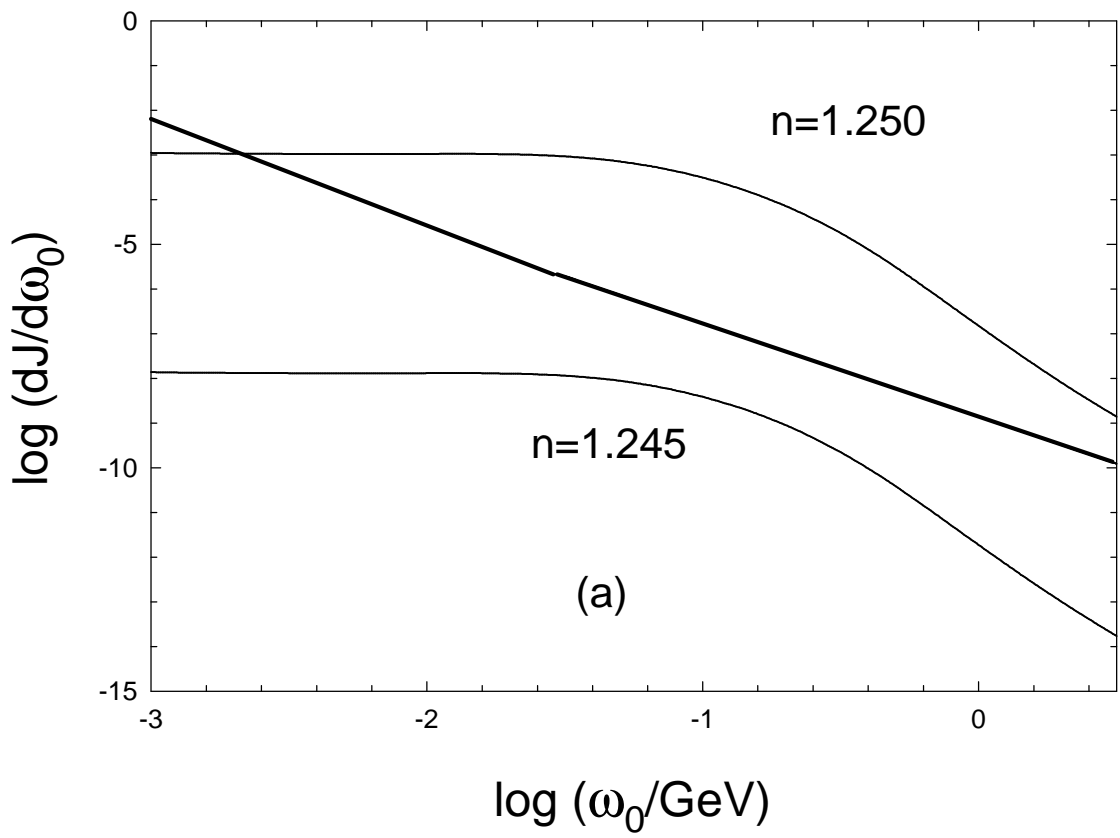
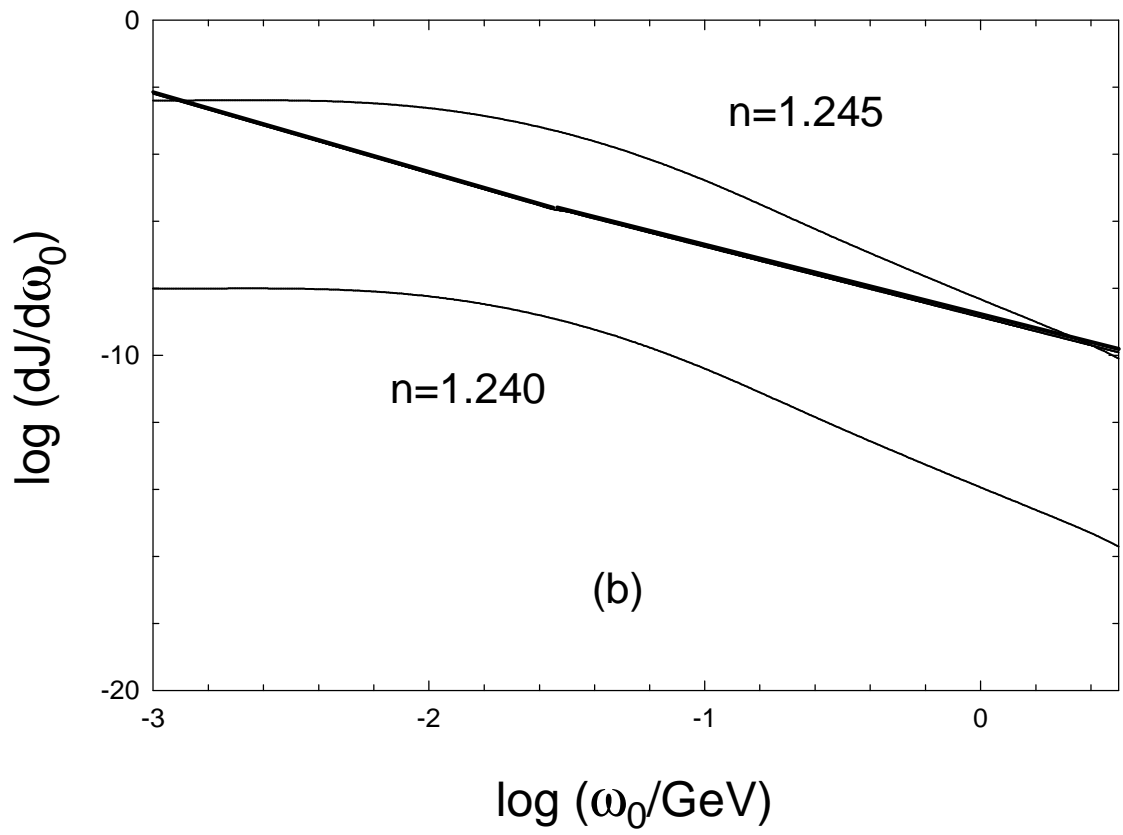
\begin{figure}
\caption{The integrated $\gamma$-ray flux from PBHs, $\frac{dJ}{d\om_{0}}$ in units of ${\rm{(cm^{2}~s~sr~MeV)^{-1}}}$, for (a) $\trh=7\times 10^{7}\gev$ corresponding to $\mbh=5\times 10^{14}\g$, and (b) $\trh=10^{8}\gev$ corresponding to $\mbh=2\times 10^{14}\g$. The bold lines are the observed flux, $\frac{dJ}{d\om_{0}}|_{\rm{obs}}$, from the COMPTEL ($1\mev<\omega_{0}<30 \mev$) and the EGRET ($30\mev <\omega_{0}<100\gev$).}
\end{figure}

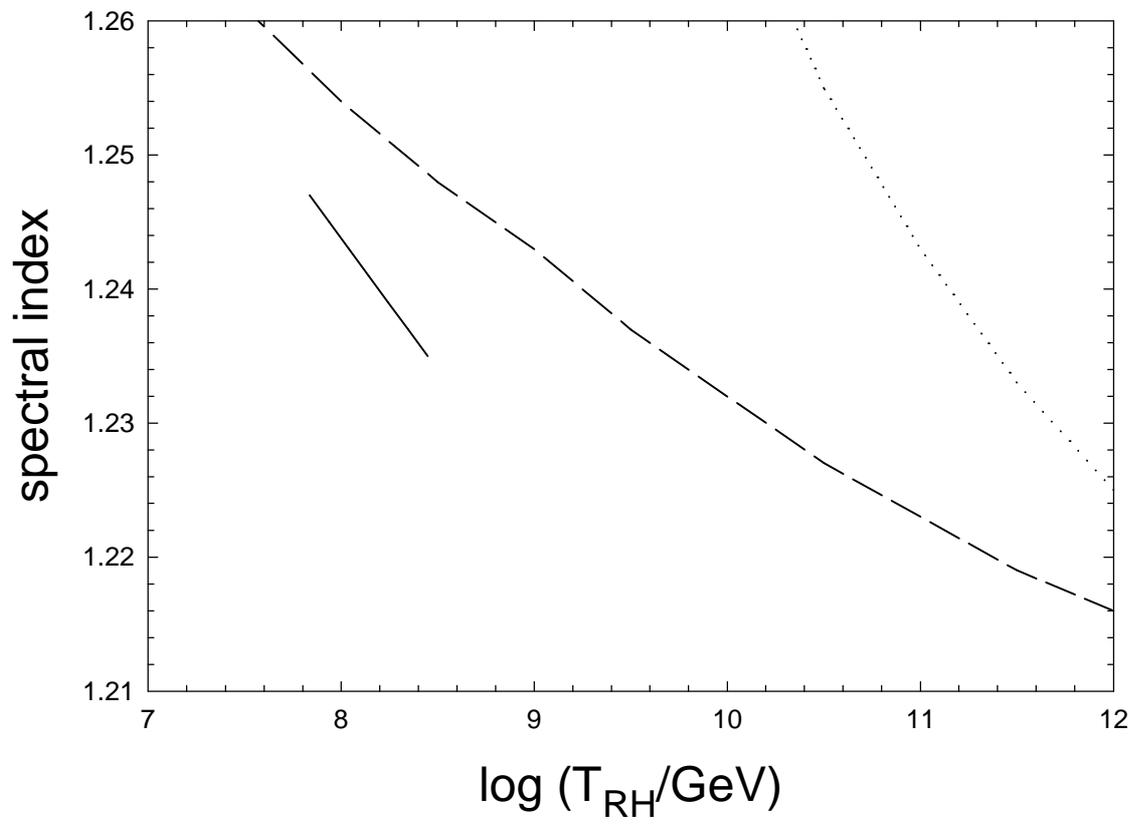
\begin{figure}
\caption{The upper limits on the spectral index. The solid line
  between $7\times 10^{7}\gev - 4\times 10^{8}\gev$ is obtained from
  the condition that the PBH $\gamma$-ray flux should not exceed the
  observed DGB flux. The dashed line is obtained from the condition
  that $\obh<1$ throughout the history of the universe (Case I). The
  dotted line is obtained from the condition that $\Omega_{\rm{relic}}<1$ (Case II).}
\end{figure}


\begin{references}

\bb{zel} Ya. B. Zel'dovich and I. D. Novikov, Astron. Zh. {\bf{43}},
758 (1966) [Sov. Astron. {\bf{10}}, 602 (1966)]; S. W. Hawking,
Mon. Not. R. Astron. Soc. {\bf{152}}, 75 (1971).  
\bb{rya} M. P. Ryan,
Astrophys. J. Lett. {\bf{177}}, L79 (1972).  
\bb{haw} S. W. Hawking,
Nature (London) {\bf{248}}, 30 (1974); Commun. Math.  Phys. {\bf{43}},
199 (1975).  
\bb{miy} S. Miyama and K. Sato, Prog. Theo. Phys. {\bf{59}},
1012 (1978); D. Lindley, \mnr {\bf{193}}, 593 (1980); T. Rothman and
R. Zatzner, Astrophys. Space. Sci. {\bf{75}}, 2291 (1981).  
\bb{car76}
B. J. Carr, \apj {\bf{206}}, 8 (1976).  
\bb{nas} P. D. Naselsky, Pisma
Astron. Zh. {\bf{4}}, 387 (1978) [Sov. Astron. Lett. {\bf{4}}, 209 (78)].
\bb{zelsta} Ya. B. Zel'dovich and A. A. Starobinsky, Pis'ma
Zh. Eksp. Teor. Fiz. {\bf{24}}, 616 (1976) [JETP Lett. {\bf{24}}, 571 (1976)].
\bb{paghaw} D. N. Page and S. W. Hawking, Astrophys. J. {\bf{206}}, 1
(1976); G. F. Chapline, Nature (London) {\bf{253}}, 251 (1975).  
\bb{maccar} J. H. MacGibbon and B. J. Carr, \apj {\bf{371}}, 447
(1991).  
\bb{hal} F. Halzen, et al., \nat {\bf{353}}, 807 (1991).
\bb{mac91} J. H. MacGibbon, \prd {\bf{44}}, 376 (1991).  
\bb{mac87}
J. H. MacGibbon, \nat {\bf{329}}, 308 (1987).  
\bb{nov} I. D. Novikov,
A. G. Polnarev, A. A. Starobinsky, and Ya. B. Zel'dovich,
Astron. Astrophys. {\bf{80}}, 104 (1979).  
\bb{cargil} B. J. Carr,
J. H. Gilbert, and J. E. Lidsey, Phys. Rev. D {\bf{50}}, 4853 (1994).
\bb{mp2} H. I. Kim and C. H. Lee, \prd {\bf{54}}, 6001 (1996).  
\bb{gre}
A. M. Green and A. R. Liddle, \prd {\bf{56}}, 6166 (1997).  
\bb{ram} R. Ramaty
and R. E. Lingenfelter, Ann. Rev. Nucl. Part. Sci. {\bf{32}}, 235 (1982); See
also, P. Sreekumar, F. W. Stecker and S. C. Kappadath,
astro-ph/9709258.  
\bb{fic}
C. E. Fichtel et al., \apj {\bf{198}}, 163 (1975).
\bb{sre}
P. Sreekumar et al., \apj 494, 523 (1998).  
\bb{pre}
J. H. MacGibbon and B. J. Carr, in preparation (1997).
\bb{mak} K. Maki, T. Mitsui and S. Orito, \prl {\bf{76}}, 3474 (1996).
\bb{wri} E. L. Wright, \apj {\bf{459}}, 487 (1996).    
\bb{gor} K. Gorski et al., \apj Lett. {\bf{464}}, L11
(1996).  
\bb{kap} S. C. Kappadath et al.,
Astron. Astrophys. suppl. {\bf{120}}, 619 (1996); S. C. Kappadath et
al,. Proc. of 4th Compton Symposium Williamburg, VA (1997). 
\bb{macweb}
J. H. MacGibbon and B. R. Webber, \prd {\bf{41}}, 3052 (1990).   
\bb{car75} B. J. Carr,
\apj {\bf{201}}, 1 (1975).  
\bb{bul} J. S. Bullock and J. R. Primack, \prd {\bf{55}}, 7423 (1997).
\bb{pag76} D. N. Page, Phys. Rev. D {\bf{14}}, 3260 (1976).  
\bb{gib} G. W. Gibbons, \com {\bf{44}}, 245 (1975).  
\bb{bir} N. D. Birrell and P. C. W. Davies, {\em
Quantum Fields in Curved Space} (Cambridge University Press, 1982).
\bb{pagpag} D. N. Page, \prd {\bf{13}}, 198 (1976); {\em ibid}. {\bf{16}}, 2402 (1977).
\bb{arn}
J. R. Arnold et al., J. Geophys. Res. {\bf{67}}, 4876 (1962); A. E. Metzger
et al., \nat {\bf{204}}, 766 (1964).  
\bb{tro} J. I. Trombka et al., \apj {\bf{212}}, 925 (1977); R. S. White et al., \apj {\bf{218}}, 920 (1977);
V. Sch\"{o}nfelder et al., \apj {\bf{240}}, 350 (1980).  
\bb{str} A. W. Strong, A. 
W. Wolfendale, and D. M. Worrall, J. Phys. A {\bf{9}}, 1553 (1976);
C. E. Fichtel, G. A. Simpson, and R. J. Thompson, \apj {\bf{222}}, 833
(1978); G. F. Bignami et al., \apj {\bf{232}}, 649 (1979).  
\bb{ste}
F. W. Stecker, D. L. Morgan, and J. Bredkamp, \prl {\bf{27}}, 1469 (1971); F. 
W. Stecker, \nat {\bf{273}}, 493 (1978).  
\bb{kri} G. D. Kribs and
I. Z. Rothstein, \prd {\bf{55}}, 4435 (1997).  
\bb{kol} E. Kolb and M. Turner, {\em The Early Universe}
(Addison-Wesley, Redwood City, CA, 1990).
\bb{zdz} A. A. Zdziarski and
R. Svensson, \apj {\bf{344}}, 551 (1989). 
\bb{nie} J. C. Niemeyer and K. Jedamzik, \prl {\bf{80}}, 5481 (1998).
\bb{hec1} A. F. Heckler, \prd {\bf{55}},
480 (1997).  
\bb{hec2} A. F. Heckler, \prl {\bf{78}}, 3430 (1997).
\end{references}
\end{document}